\documentstyle[preprint,aps]{revtex}
\begin{document}
\preprint{\vbox{ 
\hbox{SUNY-NTG-96-22} \hbox{arch-ive/9605006} }}
\draft

\title{The Effect of Nuclear Shape}
\author{S.E. Chernyshev and K.V. Shitikova
\footnote{Permanent address: Institute for Nuclear Physics, 
Moscow State University, 119899, Moscow, Russia} }
\bigskip
\date{\today} 
\maketitle
\begin{abstract}

The effect of projectile shape on cross sections and momentum 
distributions of fragments from heavy ion reactions is studied. We 
propose a new approach that implements the underlying symmetries of each 
isotope with a few parameters directly in the density. Various densities 
and their nuclear structure are then analyzed in the reactions of $^{12}$C 
and $^{11}$Li, $^{11}$Be, and $^{11}$C on a carbon target.
\end{abstract} 
\vskip 1cm

 
\narrowtext

\section{Introduction}
In a previous paper \cite{keywest} in reactions of $^{12}$C and 
$^{11}$Li projectiles on  a $^{12}$C target, it has been shown
that the total reaction cross section, inclusive fragment cross sections 
and fragment momentum distributions are sensitive to the assumed proton 
and neutron radial density distributions of the projectile. These results 
stimulated us to further investigate effects of nuclear shape of neutron 
rich isotopes in the heavy ions reactions. 
A special attention in our investigations will be given to the neutron
halo nuclei for the following reason. Recent developments of radioactive
nuclear beams enabled one to study a detailed structure of nuclei far from 
the stability line \cite{hata}. The neutron halos have been observed in 
nuclei near the neutron drip line by reaction measurements with intermediate 
and high energy radioactive nuclear beams. These nuclei are of a special 
interest in relation to shell structures near the drip line and to new 
excitation modes associated with the excess neutron on the nuclear surface.
 
In this paper we use the method of hyperspherical functions (HSF) 
\cite{kniga} to construct the densities of the light drip line nuclei.
Such nuclei typically have small binding energies and extended radial 
densities, and for them a suitable treatment of the tails of nuclear radial 
wave functions is essential. Physically such wave functions should have 
exponential tails. A main advantage of the HSF is that it provides 
a realistic radial wave functions at large distances. In addition, the 
spurious center-of-mass motion is removed from the onset and the symmetries 
are properly taken into account. In \cite{Bur94}, an attempt was made 
to provide a unified description of $^{6,7,8,9,11}$Li rather than just 
a single isotope. Instead of trying to parameterize the effective interaction 
for each isotope, a simple parametrization for all the isotopes is used. 
Furthermore, there is no inert core and all the nucleons are properly
antisymmetrized. Finally, because Jacobian coordinates are used, no problems
are encountered with the treatment of the center of mass.
The binding energy of a nucleus serves as a natural scale in this approach 
and is related to all other observables.  

In order to account for the clusterization effect more precisely in our 
calculations, we expand the HSF in the translationally 
invariant many body harmonic oscillator (HO) functions \cite{kv79} and  
implement the nuclear structure straight in the densities. We use a few 
one-body oscillator parameters in order to better reproduce the 
experimentally observed mean-square radii (rms) and density distributions.
These densities are then used in Monte Carlo simulations of 
nucleon-nucleon collisions, in which the total and inclusive 
cross sections and transverse momentum distributions of fragments are 
calculated. Thoroughly tested, this approach gives a clear picture 
of nuclear shape.   

The paper is organized as follows. In Section II, the HSF method 
is described. Next we elaborate on the details of the expansion with 
respect to the radial wave functions 
of a (3A-3)-dimensional harmonic oscillator. In Section III, we give 
a brief summary of the light isotopes and 
construct their Young diagrams. Knowing the underlying symmetries, we 
first build the "normal" $^{12}$C density and compare it with other 
approximations. We calculate the elastic form-factor and 
run a $^{12}$C + $^{12}$C simulation to see how it works in a standard 
case. Armed with the results, we proceed to the neutron rich 
$^{11}$Li and also $^{11}$Be, $^{11}$C for comparison. In order to make the  
paper more coherent, we include all the densities in Section III 
and analyze the reactions in Section IV. The results are 
in reasonable agreement with experiment and other theoretical approaches. 
Our conclusions are summarized in section V. 

\section{Method of the Hyperspherical Functions}
The HSF method involves a collective variable (hyperradius $\rho$) 
which is related to the mean-square radius of the nucleus $\rho^2 
= A<r^2>$, i.e., to the mean nuclear density. The excitations of this 
degree of freedom correspond to the monopole oscillation of the 
nucleus as a whole and thus the density can be treated as dynamic variable. 

The wave function $\Psi$ of a nucleus with $A$ nucleons is 
translationally invariant and depends on the $x_1, x_2, \ldots, x_{A-1}$ 
Jacobi coordinates, defined as 
\begin{eqnarray}
x_1 &=& \sqrt{\frac12}\big ({\bf r}_1 - {\bf r}_2 \big ) \nonumber \\
x_2 &=& \sqrt{\frac23}\bigg [ 
\frac12 ({\bf r}_1 + {\bf r}_2) - {\bf r}_3 \bigg ] \\
&\vdots& \nonumber \\
x_{A-1} &=& \sqrt{\frac{A-1}{A}}\bigg [ 
\frac1{A-1} \sum_{i=1}^{A-1} {\bf r}_i - {\bf r}_A \bigg ] \nonumber 
\end{eqnarray}
Here ${\bf r}_i$ is the coordinate of the $i-$th nucleon. Each coordinate 
is a distance between the $(i+1)$-th nucleon and and the center of mass
of the  groups of nucleons $1,2,\ldots , \, i$. 
The hyperspherical coordinates (hyperradius 
$\rho$ and $(3A-4)$ hyperspherical angles) are chosen as 
\begin{eqnarray}
x_1 &=& \rho \sin \theta_{n-1} \ldots \sin \theta_2 \sin \theta_1 
\nonumber \\
x_2 &=& \rho \sin \theta_{n-1} \ldots \sin \theta_2 \cos \theta_1 
\nonumber \\
&\vdots& \nonumber \\
x_{n-1} &=& \rho \sin \theta_{n-1} \cos \theta_{n-2} \\
x_n &=& \rho \cos \theta_{n-1} \nonumber \\
\rho^2 &=& \sum_{i=1}^{n} x_i^2  
\qquad 0 \leq \rho \leq \infty \nonumber \\
0 \leq &\theta_1 & \leq 2 \pi,  \quad  
0 \leq \theta_k \leq \pi \quad  k > 1 \nonumber \\
\nonumber 
\end{eqnarray}
for $n = 3(A-1)$-dimensional space of the Jacobi coordinates. 
For this set of coordinates, the volume element $dV$ reads 
\begin{equation}
dV = d x_1 \, d x_2 \ldots \, d x_n = \rho^{n-1} \, 
d \rho \, d\Omega
\end{equation}
where the solid angle element $d\Omega$ is 
\begin{equation}
d\Omega = \sin^{n-2} \theta_{n-1} \sin^{n-3} \theta_{n-2} \ldots 
\sin \theta_2 \; d \theta_{n-1} d \theta_{n-2} \ldots d \theta_1
\end{equation}
The Laplacian is given by 
\begin{eqnarray}
\Delta_n &=& \sum_n \frac{\partial^2}{\partial x^2_n} = 
\frac1{\rho^{n-1}} \frac{\partial}{\partial \rho} 
\left ( \rho^{n-1} \frac{\partial}{\partial \rho} \right )  
+ \frac1{\rho^2} \Delta_{\Omega_n} \\
\Delta_{\Omega_n} &=& \frac1{\sin^{n-2} \theta_{n-1}}
\frac{\partial}{\partial \theta_{n-1}}
\, \left ( \sin^{n-2} \theta_{n-1} 
\frac{\partial}{\partial \theta_{n-1}} \right ) 
+ \frac1{\sin^2 \theta_{n-1}} \Delta_{\Omega_{n-1}} \nonumber 
\end{eqnarray}

The hyperspherical functions, or $K-$harmonics, 
are the eigenfunctions of the angular part of the Laplacian 
\begin{equation}
\Delta_{\Omega_n} Y_{K\gamma}(\theta_i)=-K(K+n-2)Y_{K\gamma}(\theta _i).
\label{ang}
\end{equation}
The value of $K$ is the analog of the angular momentum at $n=3$ and is called
the global momentum. The subscript $\gamma$ denotes all the quantum numbers 
necessary to enumerate the various degenerate states of Eq. (\ref{ang}). 
For $\gamma$ it is expedient to use the Young diagram 
$[f]$ and the Yamanouchi symbol $(r)$, characterizing the $K-$harmonics 
properties relative to the $A-$nucleon permutations, and $LST$ to designate 
the orbital momentum, spin and isospin of this state. 

The wave function of a nucleus is then expressed in the form of an expansion 
in the $K-$harmonic polynomials \cite{sim} 
\begin{eqnarray}
\Psi (1,2...A) &=& \rho^{-(3A-4)/2} \sum_{K\gamma}
\chi_{K\gamma}(\rho) Y_{K\gamma}(\theta) \nonumber \\
\int_0^\infty  \chi_{K\gamma}^2 (\rho) d \rho &=& 1 \qquad 
\gamma = [f] r LST \label{expan}
\end{eqnarray}
The Hamiltonian reads 
\begin{equation}
{\cal H} = - \frac{\hbar^2}{2m} \, 
\frac1{\rho^{3A-4}} \frac{\partial}{\partial \rho} 
\left ( \rho^{3A-4} \frac{\partial}{\partial \rho} \right )  
- \frac{\hbar^2}{2m} \frac{\Delta_{\Omega}}{\rho^2} + V 
\end{equation}
Here $m$ is the mass of a nucleon. 
The Schroedinger equation 
for the radial function $\chi_{K\gamma}(\rho)$ can be written as 
\begin{eqnarray}
\left \{ \frac{d^2}{d\rho^2} - \frac{L_K(L_K+1)}{\rho^2}
- \frac {2m}{\hbar^2} \big [ {\cal E} + W^{K\gamma}_{K\gamma}(\rho)
\big ] \right \} \, \chi_{K\gamma}(\rho)
= \frac {2m}{\hbar^2} \sum_{K^\prime\gamma^\prime \neq K\gamma}
W^{K^\prime\gamma^\prime}_{K\gamma}(\rho)
\chi_{K^\prime\gamma^\prime}(\rho) \label{schrod}
\end{eqnarray}
where $L_K = K + (3A-6)/2$, and $W^{K^\prime\gamma^\prime}_{K\gamma}(\rho)$ 
are the matrix elements of the nucleon-nucleon interaction 
\begin{equation}
V = \sum_{i<j}^A V(r_{ij}) \qquad V(r_{ij}) = f(r_{ij}) W_{\sigma\tau}
\end{equation}
which may be expressed through the two-body fractional parentage 
coefficients in the form 
\begin{eqnarray}
W^{K^\prime\gamma^\prime}_{K\gamma}(\rho) &=& 
<AK[f] r LST M_L M_S M_T | \hat V | AK^\prime [f] r^\prime 
L^\prime S^\prime T^\prime M_{L^\prime}  M_{S^\prime}  M_{T^\prime} > 
\nonumber \\
&=& \frac{A(A-1)}{2} \sum_{K_2, \gamma_2} 
<AK[f] r LST | (A-2) K_2 [f] r_2 
L_2 S_2 T_2 \Lambda(L^{\prime \prime}  K^{\prime \prime});  L_0 S_0 T_0 > 
\nonumber \\ &\times&
<S_0 T_0 | W_{\sigma\tau}|S_0 T_0 > \, 
R^{K K^\prime}_{K^{\prime \prime} L_0} (\rho) 
\nonumber \\ &\times&
<(A-2) K_2 [f] r_2 L_2 S_2 T_2 \Lambda(L^{\prime \prime}  
K^{\prime \prime});  L_0 S_0 T_0 | 
AK^\prime [f] r^\prime L^\prime S^\prime T^\prime > 
\end{eqnarray}
Here 
\begin{eqnarray}  
R^{K K^\prime}_{K^{\prime \prime} L_0} (\rho) &=& 
\int d \theta_1 \sin^{3A-7} \theta_1 \, \cos^2  \theta_1 \; 
{\cal N}_{K K^{\prime \prime} L_0}
{\cal N}_{K^\prime  K^{\prime \prime} L_0} 
\nonumber \\ &\times&
f(\rho \cos \theta_1 ) \sin^{2K^{\prime \prime}}\, \cos^{2L_0} \theta_1 \; 
P_{K - K^{\prime \prime} - L_0}^{K^{\prime \prime} + (3A-6)/2 - 1, 
L_0 + 1/2}(\cos 2\theta_1 )
\nonumber \\ &\times&
P_{K^\prime - K^{\prime \prime} - L_0}
^{K^{\prime \prime} + (3A-6)/2 - 1, L_0 + 1/2} (\cos 2\theta_1 )
\end{eqnarray}
$<S_0 T_0 | W_{\sigma\tau}|S_0 T_0 > $ is the spin-isospin part of the 
interaction matrix element. The collective potential 
$W^{K^\prime\gamma^\prime}_{K\gamma}(\rho)$ is given by integrals 
containing two-body matrix elements together with the $K-$harmonics 
fractional parentage coefficients. In the previous investigations 
\cite{ss77obsor} it was 
shown that with the HSF one may use the same formulas for the 
fractional parentage coefficients which were obtained earlier with the 
translationally invariant shell model. In this case the Talmi-Moshinsky 
coefficients must be replaced by the Raynal-Revai coefficients, 
6j-symbols must be added for the over-binding of the global 
momentum $K$, and additional phase multipliers are inserted. Their use 
has led to a significant improvement of the computations \cite{dji89}. 
The calculation is simplified if only the first 
few terms in the expansion (\ref{expan}), i.e., $K=K_{min}$ and 
$K=K_{min}+1$ are taken into account. Usually one adopts 
the $K_{min}$ approximation in which all 
values of K greater than $K_{min}=A-4$ ($A\leq 16$) 
are neglected. The success of 
this approximation lies in the fact that the centrifugal barrier reduces 
the contributions of configurations with $K$ greater than 
$K_{min}$ in the equations determining the hyperspherical wave functions. 

\subsection{Expansion}

The system (\ref{schrod}) has an analytical solution for the harmonic 
oscillator potential $V = m \omega^2 \rho^2 /2$ \cite{ss77obsor}. 
The energy eigenvalues 
are given by 
\begin{equation}
{\cal E}_\nu = (2\nu + K + \frac n2 ) \, \hbar\omega 
\end{equation}
with the eigenfunctions 
\begin{equation}
{\cal R}_{\nu K} (x) =\sqrt{\frac{2 \nu ! } {\Gamma(\nu + K + \frac n2)}}
\, e^{-\frac 12 x^2} x^K L_\nu^{K+ \frac 12 (n-2)} (x^2) 
\label{eigen}
\end{equation}
where we have introduced a dimensionless parameter $x = \rho/r_0$
with $r_0^2 = \hbar/m\omega$ and the associated Laguerre 
polynomials are defined as
\begin{equation}
L_n^\alpha (x) = \sum_{k=0}^{n} (-1)^k \frac{(n+\alpha)!}
{(n-k)!\,  (k+\alpha)! \, k!\, } \; x^k
\end{equation}
Let us briefly examine critical regimes for the Schroedinger 
equation (\ref{schrod}). As $\rho \rightarrow 0$, it is reduced to
\begin{equation}
\frac{d^2}{d\rho^2} \chi(\rho) =  \frac{L_K(L_K+1)}{\rho^2} \chi(\rho)
\end{equation}
with the trivial eigenfunction 
\begin{equation}
\chi(\rho) \sim \rho^{L_K+1} 
\end{equation}
On the other hand, as $\rho \rightarrow \infty$, the Eq. (\ref{schrod}) 
reads   
\begin{equation}
\frac{d^2}{d\rho^2} \chi(\rho) =  
\frac{2mE}{\hbar^2} \chi(\rho)
\end{equation}
and the eigenfunction has the following asymptotical behavior 
\begin{equation}
\chi(\rho) \sim \exp{\left( - \sqrt{\frac{2mE}{\hbar^2}} \; \rho \right ) } 
\label{asym} 
\end{equation}
as was mentioned in the Introduction. We will return to this important 
point in next the section.

Now we can express radial functions $\chi_{K}^{\gamma}(\rho)$ 
in terms of the $3(A-1)$-dimensional harmonic oscillator, as shown on 
Figure 1 for the ${}^{12}$C nucleus,    
\begin{equation}
\chi_{K}^{\gamma}(\rho) = \sum_\nu C^\gamma_{\nu K} 
R_{\nu K} (\rho) \label{sumho}
\end{equation}
and the expansion coefficients are given by
\begin{eqnarray}  
C^\gamma_{\nu K} &=& \int \chi_{K}^{\gamma}(\rho) 
R_{\nu K} (\rho) \; \rho^{\frac12(3A-4)} \, d\rho \label{overlap} \\
C_0 &\simeq& 0.984, \qquad  C_1 \simeq -0.061, \nonumber \\
C_2 &\simeq& 0.158, \qquad  C_3 \simeq 0.055. \nonumber 
\end{eqnarray}  
with the oscillator frequency $\hbar\omega$ determined so that the maximum 
of the lowest oscillator function $R_{0K}(\rho)$ coincides with the maximum 
of the lowest hyperspherical function $\chi_K^0 (\rho)$. We will 
deal only with the ground state wave function and use 
$C_{\nu K_{min} }^0 = C_\nu$ for short.  

It is striking that the overlap integral of the ground state radial 
function $\chi_K^0(\rho)$ with the lowest oscillator function 
$R_{0K}(\rho)$ is $98\%$. 
At the same time, the contribution to $\chi_K^0(\rho)$ of the two-quantum 
oscillator excitation is exceptionally small. A very small value of the 
coefficient $C_1$ is not fortuitous but conforms to the theorem \cite{50}, 
which states that if the oscillator frequency $\hbar\omega$ is chosen to 
make the overlap integral of $R_{0K}(\rho)$ and $\chi_K^0(\rho)$ maximal, 
then the overlap integral of $\chi_K^0(\rho)$ with $R_{1K}(\rho)$ is 
strictly zero. 

As we have already noted, in choosing $\hbar\omega$ we did not achieve 
a maximum value of the $\chi_K^0(\rho)$, $R_{0K}(\rho)$ overlap, 
but assured only a coincidence of the maxima of these functions. 
However, if such a condition is satisfied, the overlap integral of the 
functions is fairly near the maximum, and the coefficient $C_1$ 
is very small, although nonvanishing. 
This circumstance casts light on the origin of close agreement between the 
results of the translationally invariant shell model and the HSF method 
for the ground state. Since the function $R_{1K}(\rho)$ cannot be mixed with 
$R_{0K}(\rho)$, the shell function in the HSF method is improved by admixture 
of states with oscillator energy $4\hbar\omega$ and higher. Because of the 
large difference between the energies of these states and the ground state,  
this admixture is small.

Then one can derive the binding energy as a function of the interaction 
matrix elements
\begin{eqnarray}
E_{\nu K} &=& \frac 12 \hbar\omega \; \big ( \; 2 \nu + \frac{5A-11}{2} \big ) 
\nonumber \\ &+& 
\frac 12 \hbar\omega \left [ \, \frac {C_{\nu+1}}{C_\nu}\sqrt{
(\nu + 1) \big (\nu + \frac{5A-11}{2} \big ) } + 
\frac {C_{\nu-1}}{C_\nu}\sqrt{\nu \big (\nu + \frac{5A-13}{2} \big ) } 
\; \right ] \nonumber \\
&+& \sum_{\mu} \frac {C_\mu}{C_\nu} \, \int R_{\nu K} (\rho) 
W^{K \mu}_{K\nu} (\rho) 
R_{\mu K} (\rho) \rho^{3A-4} d\rho \label{bind}
\end{eqnarray}
The rms of a nucleus is given by
\begin{eqnarray}
<\bar r^2>_\nu &=& \frac{\hbar}{m\omega} \frac 1A\, 
\left [ \, \sum_\nu C_{\nu}C_\nu (2 \nu + \frac{5A-11}{2} \big ) -
2 \sum_\nu  C_\nu C_{\nu+1}
\sqrt{ (\nu + 1) \big (\nu + \frac{5A-11}{2} \big ) } \, \right ]
\label{rmso}
\end{eqnarray}

\section{Symmetries of neutron rich isotopes} 
Using the Young-Yamanouchi classification scheme \cite{jahn}, 
which stems from the orthogonal representation of a given 
permutation group, and with the knowledge of the total spin and isospin 
of the nucleus, we determine the symmetry with a corresponding 
Young diagram $[f]$ of each isotope.  In Table I results of the 
classification of ${}^{6-13}$Li, ${}^{6-12}$Be, ${}^{7-13}$B (for 
completeness), and ${}^{8-14}$C nuclei are collected. The Lithium sequence, 
illustrated in Figure 2, reflects the strong excitation  in the isospin. 
The neutron super-rich ${}^{13}$Li obtains an extra pair of halos in the 
same manner as ${}^{11}$Li results from ${}^{9}$Li by adding two halo 
neutrons.  In the brackets we have put two possible versions of a 
transitory ${}^{10}$Li state. 

Note that while for the most part these sequences are rather 
straightforward, obtained by a simple adding of a neutron to the 
p-shell, some isotopes like ${}^{11}$Be do stand out. There has been 
a great deal of interest in the effect of clusterization, and our 
scheme seems to agree with a recent study of the multi-cluster 
structure \cite{nucho}, that a simple description in terms of one 
Young diagram may not be adequate for some isotopes. 

Having thus established the symmetry requirements on the neutron content of 
different isotopes, we proceed to construct the density distributions for 
the light nuclei. Originally, this classification was applied to the 
construction of the HSF wave functions \cite{Bur94}. 
In Table II we reproduce their results for the binding energy and rms 
for ${}^{6-11}$Li and $^{11}$C isotopes in comparison with the 
experimental data. It can be seen that the variation 
in the binding energy as a function of the mass number and rms 
for the Lithium isotopes is qualitatively reproduced. 
A comparison of $^{11}$Li and $^{11}$C shows a very strong nuclear 
structure effect on the binding energy and rms. 

If one would like to look at the more detailed description of the nuclear 
structure which comes with the wave function and exhibits the nuclear shape 
through the nuclear densities, one may use the information that comes 
from measurements of the total reaction cross sections and the inclusive 
cross sections for fragments in which the projectile has lost 
one neutron or proton. In Ref. \cite{keywest} it has been shown that 
both the cross sections and the fragment momentum distributions lead 
to the conclusion that the HSF density distributions in the $K_{min}$ 
approximation extend too far for protons and not far enough for neutrons. 
The inclusion of several more terms in the expansion (\ref{expan}) will 
give an extension of the asymptotic density to larger radii. This will give 
better agreement between the HSF method and the experimental data. 
For this reason we propose another approach in which the 
clusterization effect is taken into account directly in the density. 

\subsection{Density}
From the formulae in section II one derives the density 
of nuclear matter through 
\begin{equation}
n_{J^\pi J^{\prime \pi^\prime}}  ({\bf r}) = 
\bigg \langle J^\pi \bigg | \sum_K \delta ({\bf r} - {\bf r}_K) \bigg | 
J^{\prime \pi^\prime} \bigg \rangle 
\end{equation}
with 
\begin{equation}
n_{J^\pi J^{\prime \pi^\prime}} ({\bf r}) = 
\sum_{\lambda \mu} 
n^\lambda_{J^\pi J^{\prime \pi^\prime}} (r) Y_{\lambda \mu} (\theta, \phi)
\end{equation}
and the radial component is expressed as 
\begin{equation}
n^\lambda_{J^\pi J^{\prime \pi^\prime}} ({\bf r}) = 
\bigg \langle J^\pi \bigg | \sum_K {\bf r}_k^{-2} 
\delta ({\bf r} - {\bf r}_K) 
Y_{\lambda \mu} (\theta, \phi) \bigg | 
J^{\prime \pi^\prime} \bigg \rangle 
\end{equation}

The radial density distribution reads ($A\leq 16$) \cite{dji89} 
\begin{eqnarray}
n_{ij}(r) 
&=& \frac {16}{\sqrt{\pi}} \frac
{\Gamma\left (\frac12(5A-11)\right )}{\Gamma\left (\frac12(5A-14)\right )}
\int_r^\infty \, 
\frac{ \left( \rho^2 -r^2 \right )^{\frac12(5A-16)} }{\rho^{5A-13} }
\, \chi_i (\rho)  \chi_j (\rho) \; d\rho \nonumber \\
&+& \frac 83 \frac{A-4}{\sqrt{\pi}} \frac
{\Gamma\left (\frac12(5A-11)\right )}{\Gamma\left (\frac12(5A-16)\right )}
\int_r^\infty \, 
\frac{ r^2 \left( \rho^2 - r^2 \right )^{\frac12(5A-18)} }{\rho^{5A-13} }
\, \chi_i (\rho)  \chi_j (\rho) \; d\rho \label{dense}
\end{eqnarray}
with the s-shell (first line) and p-shell densities explicitly shown. 
We also treat the halo neutrons $N_h$ separately from the p-shell nucleons.
The rms of each shell is calculated as 
\begin{equation}
<r_{ii}^2>_{shell} = \frac { \int n_{ii}^{shell} (r) r^4 dr}
{\int n_{ii}^{shell} (r) r^2 dr}
\end{equation}
with the density normalization 
\begin{eqnarray}
4 \pi \int_0^\infty n_{ii}^{(s,p,h)}
(r) r^2 dr = (\, 4 , \; A-4 - N_h , \; N_h\, ) 
\end{eqnarray}

Before calculating the densities for the neutron rich isotopes, 
we would like to check on a standard nucleus such as $^{12}$C, 
for which there exist a few other density approximations.

\begin{itemize} 
\item In Ref. \cite{bogd} the exact HSF density for 
${}^{4}$He, ${}^{6}$Li, ${}^{12}$C, and ${}^{16}$O was 
approximated by 
\begin{equation}
n (r)= \sum_{i=0}^k \, a_i \, \frac{r^{2i}}{r_i^{2i} } \, 
e^{- r^2/r_i^2 } 
\end{equation}  
motivated by the analogy with the harmonic oscillator. 
Using different oscillator parameters $a_i, \; r_i$, one could easily 
fit various experimental observables, such as the rms and so on. 
For the ground state of ${}^{12}$C, 
it was enough to have the first two terms
\begin{equation}
n (r)= a_s e^{- r^2/r_s^2 } + a_p \,
\frac{r^2}{r_p^2} \, e^{- r^2/r_p^2 } \label{hoho} 
\end{equation}  
with $r_s= 2.0\, $fm, $r_p= 1.7\, $fm and $a_s= 0.718348$, $a_p= 0.957798$ 
fixed to the normalization of 4 s-shell and 8 p-shell nucleons. 

\item Harvey used a phenomenological approach \cite{Har94} to obtain the 
densities for ${}^{12}$C and ${}^{11}$Li. 
Since the Monte Carlo simulations require a separate density distribution
for nucleons with different binding energies, a harmonic oscillator model 
was used for projectiles and a standard Fermi shape for the 
${}^{12}$C target. For example, for $^{11}$Li the $s-$shell parameter 
was chosen to be $1.619\, $fm for protons and neutrons, and for the 
$p$-shell $2.0\, $fm for protons and $2.1\, $fm for neutrons. 
It was also assumed that there was one proton and four neutrons in 
the $p_{3/2}$-shell. The outer two neutrons were assumed to have an
exponential density distribution beyond a radius of $2.5 \, $fm
\begin{equation}
n (r)=0.3323\frac{e^{-2r/L}}{r^2}  \label{harvey}
\end{equation}  
with $L$=3.7 fm. As we mentioned above, the HSF functions have 
an exponential asymptotics (\ref{asym}) and thus become very attractive 
from a phenomenological point of view. The normalization, 0.3323, came 
from the requirement that the volume integral contain just two neutrons.
The values of the harmonic oscillator parameters and of $L$ were chosen 
to fit the experimental reaction cross section and the $^{9}$Li, $^{8}$He 
inclusive cross sections for the reaction $^{11}$Li+C at 790 MeV/A.

\item Returning to the HSF method, one does not have as much freedom 
in choosing various parameters. It is very important that in the HSF 
one obtained a self-adjusted system in which the parameters of the 
interaction determine the size of the system. However, working in the 
$K_{min}$ approximation, one needs to improve the neutron density 
asymptotics. Having derived the explicit formula (\ref{dense}) for 
the density distribution and knowing the symmetry requirements on the 
neutron content of different isotopes, we are now in the position to do 
it directly in the density. 
\end{itemize}

Since the wave function characterizes the nucleus as a whole, 
the binding energy E$_{0 K}$, cf. Eq. (\ref{bind}), 
sets its scale $r_A = \hbar/m\omega$. It also serves as a natural 
parameter for the s-shell, the core of a nucleus. The p-shell basically 
carries all the information about a given isotope, and its 
scale will be chosen so that experimental rms is reproduced. 
In accordance with our group-theoretical framework, another parameter 
will be used for the halo neutrons, still fitting to the experimental rms. 
  
For example, the p-shell density then reads
\begin{eqnarray}
n_{ij}^p(r) = \frac 83 \frac{A-4}{\sqrt{\pi}} \frac
{\Gamma\left (\frac12(5A-11)\right )}{\Gamma\left (\frac12(5A-16)\right )}
\int_r^\infty \, 
\frac{ r^2 \left( \rho_p^2 - r^2 \right )^{\frac12(5A-18)} }{\rho_p^{5A-13} }
\, \chi_i (\rho_A)  \chi_j (\rho_A) \; d\rho_p 
\end{eqnarray}
with scaling
\begin{eqnarray}
\rho_p \rightarrow \frac{\rho}{r_p} \qquad  
\rho_A \rightarrow \frac{\rho}{r_A}
\end{eqnarray}

Table III shows the parameters for these densities, the rms for each 
shell and of the nucleus as a whole. They reproduce the experimental 
radii of Tanihata, et al. \cite{hata,t88}. 
In Figure 3 we plot the densities for ${}^{12}$C and ${}^{11}$Li. 
The ${}^{11}$Li one is in reasonable agreement with the shaded region 
of the experimental density \cite{hatad}.   

\subsection{Charged form-factors}

Having obtained the total densities, one can quickly probe its 
charge distribution by looking at the elastic $e^-$ scattering off 
that nucleus, or its charged form-factor. For a nucleus with $Z$ 
protons it may be written in the Born approximation as 
\begin{equation}
F_{ch}(q) = f_p(q) \; f_{CM}(q) \; \frac 1Z \, F_{z}(q) 
\end{equation}
with the elastic point-proton form-factor $F_{z}(q)$ defined as 
\begin{equation}
F_{z}(q) = 4 \pi \int_0^\infty n_{z}(r) \frac {\sin (qr)}{qr} r^2 dr
\end{equation}
where $n_{z}(r)$ is the proton part of the total density.   
The correction for the center of mass motion $f_{CM}(q)$ 
is not required in our approach since it is factorized from the very 
beginning (cf. Eq. \ref{schrod}). 
$f_p(q)$ is the correction for the finite proton size, 
which we take to be 
\begin{equation}
f_p(q) = [\, 1 + q^2/q_0^2\, ]^{-2} 
\end{equation}
with $q_0 = 4.33 \, $fm${}^{-1}$ corresponding to $r_{rms} =0.8\, $fm 
for the proton. 

In Figure 4 we plot the charged form-factors for 
${}^{6}$Li and ${}^{12}$C with the data points taken from \cite{c12,li6}, 
where for the momentum axis a standard $q_{eff}$ is used 
\begin{equation}
q_{eff} = q\; [\, 1 + \frac{ 4Z\alpha \hbar c}{3r_{rms} E_0 } \; ] 
\end{equation}
Note, that the elastic rms radii for these nuclei are different from the 
reaction ones \cite{hata}, and we have slightly modified our parameters so 
that their densities reproduce the observed $r_{rms}({}^{6}{\rm Li}) = 
2.56\, $fm and $r_{rms}({}^{12}{\rm C})= 2.48\, $fm. 
Our results match well the minima of the experimental form-factors and 
qualitatively follow their shape. They also compare 
favorably with the full HSF calculation \cite{Bur94,Bur81}.

Having determined all the parameters in the density, we are 
ready to analyze the reactions. 

\section{Monte Carlo and Nuclear Shape}

The Monte Carlo method allows the calculation of cross sections and 
momentum distributions of primary fragments arising from 
nucleon-nucleon collisions between a target and a projectile at 
energies above 200 MeV/nucleon. The primary fragments may 
subsequently decay to the final observed nuclei.

This method has been shown \cite{Har85} to give results in good agreement 
with experiment for the total reaction cross section, fragment inclusive 
cross sections, fragment exclusive (coincidence channel) yields, fragment 
momenta, both inclusive and in different coincidence channels. 
It is also possible to calculate the momentum distributions of nucleons that 
are either knocked out of the projectile or come from the subsequent decay. 
Thus, unlike many theoretical techniques, the Monte Carlo decay method 
permits the comparison of calculated and experimental values for 
essentially all the measured quantities and allows to test radial density 
distributions derived from various theoretical models. The purpose of 
the present work is therefore to compare experimental values with those 
obtained from the HSF density distributions with the aim to get the maximum 
of information about nuclear shape.

\subsection{${}^{12}$C + ${}^{12}$C} 

When one or two neutrons are lost in a collision between the projectile 
with a diffuse neutron rich surface and a target nucleus, 
the transverse momentum distribution of the fragments thus produced can be fit 
with a double Gaussian curve in which a narrow component is superimposed on a 
much broader one \cite{Har94}. This phenomenon is also observed for 
fragments from the normal projectile $^{12}$C. 
The shape of the fragment momentum distribution is sensitive to the 
assumed projectile surface diffusivity. As the surface diffusivity increases, 
the width of the broad component decreases. For the sharp cutoff shape, 
the distribution has no narrow component. 

The study of the nuclear structure effect in ${}^{12}$C + ${}^{12}$C 
scattering, using the optical model \cite{scat92} and the calculation of 
the total reaction cross sections in the Glauber model \cite{tot91} 
convincingly show that the total reaction cross sections 
are sensitive to the form of the density. With that in mind, we have 
conducted the Monte Carlo simulations with three different densities:
minimal $n(R_{0 K})$, obtained with $R_{0 K}$ instead of the full 
function $\chi$ of Eq. (\ref{sumho}), $n(\chi)$ and the approximation 
of Eq. (\ref{hoho}). The experimental data and our results are collected 
in Table IV. As expected, the total cross sections did not vary much, 
since we had the same $r_{rms}$ for each density distribution. 

The tail effect is deduced from the momentum distribution of $^{11}$C 
fragments. In a fit with the double Gaussian curve, 
the narrow component reflects the behavior of 
the tail of the density and thus we see that having all four 
functions of the Eq. (\ref{sumho}) decomposition improves the resulting 
width of both narrow and broad components as compared to 
the minimal $n(R_{0 K})$ density. 

In Figure 5 we fit the transverse momentum distribution of 
the $^{11}$C fragments from our $n(\chi)$ density with one 
and two Gaussian curves. One Gaussian fit is quite reasonable and 
gives a very close $\sigma_B=103$ to the experimental $\sigma_B=105$. 
Overall, our numbers match well with the experimental data and 
phenomenological results of Ref. \cite {Har94}. 

Having thus convinced ourselves that this approach works well 
for the standard nucleus, we turn to the reactions with the 
neutron rich isotopes as projectiles on the carbon target. 

\subsection{${}^{11}$Li + ${}^{12}$C}

In analyzing this reaction, we emphasize the halo effect. 
The momentum distribution of $^{9}$Li fragments from $^{11}$Li 
+ C at 790 MeV/c has been experimentally studied \cite{Kob88}. 
It has broad and narrow Gaussian components with widths of 80 and 21 
MeV/c and a ratio of areas B/N of 1.5. Since the shape of the momentum 
distribution is sensitive to the projectile surface diffusivity,
we have used two different HSF densities with a very loose pair of 
halo neutrons $n(^{11}$Li$_{h})$ and another one $n(^{11}$Li$_{p})$ by 
placing them relatively close to the p-shell. The results for 
$n(^{11}$Li$_{h})$ are shown in Figure 6 and the rest is summarized 
in Table V. While the width of the broad component is well reproduced,
the narrow component is too wide, especially for $n(^{11}$Li$_{p})$.  
This result is consistent with the absence of a long neutron `tail' 
at large radii and low local neutron densities which give a small value for 
the local Fermi momentum. A short tail also manifests itself in the somewhat 
low total cross sections, which drops even more for $n(^{11}$Li$_{p})$. This 
indicates that the halo pair gives a major contribution to the neutron tail 
and has an unusually large rms radius. We would like to point out that our 
result $\sqrt{<r^2>_o} \sim 5.23\, $fm for an outer pair of halo neutrons 
for $n(^{11}$Li$_{h})$ is consistent with experimental 
$4.8 \pm 0.8\, $fm \cite{hatad} and $5.1^{+0.6}_{-0.9}\, $fm \cite{expli}. 

At last, we would like to analyze the structure effect by comparing 
${}^{11}$Li, ${}^{11}$Be, and ${}^{11}$C, which have the same number of 
nucleons but different Young diagrams. From ${}^{11}$C with $[f]= [443]$ 
and isospin $T=1/2$, one sees the increase in the isospin in ${}^{11}$Be 
with $[f]=[4331]$, $T=3/2$ to ${}^{11}$Li with $[f]= [4322]$, $T=5/2$. 
Their cross sections, see Table VI, vividly show this change in their 
nuclear structure. Starting with $\sigma_{{}^{11}C}=880\, $mb, the cross 
section jumps to $\sigma_R = 946\,$mb for ${}^{11}$Be, indicating some sort 
of a halo neutron tail and culminates in $\sigma_{{}^{11}Li} = 994\,$mb with 
two halo neutrons. It unambiguously verifies that our approach correctly 
implements the underlying symmetries of each isotope and works well with a 
judicious choice of parameters directly in the density level, bypassing 
the parameters of the nucleon-nucleon interaction. 

\section{Conclusion}
The hyperspherical functions method provides a convenient basis for the 
description of the neutron rich isotopes. The underlying symmetry of each 
isotope exhibits a simple structure and reproduces the clusterization effect. 
Studies of the reaction cross section and the inclusive cross sections for
fragments in which the projectile has lost one neutron or one proton give a 
powerful method for investigating the neutron, proton and total densities in
the surface of the projectile. Here we have proposed a new approach 
which takes into account the clusterization effect directly 
to the densities of the drip line nuclei. 

\acknowledgements

We would like to express our deep gratitude to Prof. B.G. Harvey for many 
useful discussions and great help with the Monte Carlo part of the calculations.
One of the authors gratefully acknowledges the hospitality of the 
Nuclear Theory Group at Stony Brook. This work was supported in part by 
the Department of Energy under Grant No.\, DE-FG02-88ER40388.

\begin{table}
\caption{Classification of the isotopes} 
\begin{center}
\begin{tabular}{|c|c|c|c|}  
$^A$Z     & $J^\pi$  & T & [f] \\
\hline 
$^{6}$Li  & $1^+$    & 0   & [42]    \\[-.2cm]
$^{7}$Li  & $3/2^-$  & 1/2 & [43] [421]  \\[-.2cm]
$^{8}$Li  & $2^+$    & 1   & [431] [422]  \\[-.2cm] 
$^{9}$Li  & $3/2^-$  & 3/2 & [432] [4221] \\[-.2cm]
($^{10}$Li) & $(1^+)$ & 2 & [4222] \\[-.2cm]
($^{10}$Li) & $(2^+)$ & 2 & [4321] \\[-.2cm]
$^{11}$Li & $3/2^-$  & 5/2 & [4322] \\[-.2cm]
$^{13}$Li & $3/2^-$  & 7/2 & [43222] \\
\hline 
$^{6}$Be  & $0^+$    & 1   & [42]   \\[-.2cm]
$^{7}$Be  & $3/2^-$  & 1/2 & [43] [421]  \\[-.2cm]
$^{8}$Be  & $0^+$    & 0   & [44] [422] \\[-.2cm]
$^{9}$Be  & $3/2^-$  & 1/2 & [441] [432] [4221] \\[-.2cm] 
$^{10}$Be & $0^+$    & 1   & [442] [4321] [4222] \\[-.2cm] 
$^{11}$Be & $1/2^+$  & 3/2 & [4331] \\[-.2cm]
$^{12}$Be & $0^+$    & 2   & [4431] [4422] \\
\hline 
$^{7}$B  & $3/2^-$ & 3/2 & [421]  \\[-.2cm]
$^{8}$B  & $2^+$   &  1  & [431] [422]  \\[-.2cm]
$^{9}$B  & $3/2^-$ & 1/2 & [441] [432] [4221] \\[-.2cm]
$^{10}$B & $3^+$   & 0   & [442] [4321] [4222] \\[-.2cm]
$^{11}$B & $3/2^-$ & 1/2 & [443] [4421] [4322] \\[-.2cm]
$^{12}$B & $1^+$   & 1   & [4431] [4422] \\[-.2cm]
$^{13}$B & $3/2^-$ & 3/2 & [4432] \\
\hline 
$^{8}$C   & $0^+$    & 2   & [422]  \\[-.2cm]
$^{9}$C   & $3/2^-$  & 3/2 & [432] [4221]  \\[-.2cm]
$^{10}$C  & $0^+$    & 1   & [442] [4321] [4222]  \\[-.2cm]
$^{11}$C  & $3/2^-$  & 1/2 & [443] [4421] [4322] \\[-.2cm] 
$^{12}$C  & $0^+$    & 0   & [444] [4422] \\[-.2cm] 
$^{13}$C  & $1/2^-$  & 1/2 & [4441] [4432] \\[-.2cm] 
$^{14}$C  & $0^+$    & 1   & [4442] 
\end{tabular}
\end{center}
\end{table}

\begin{table}
\caption{Calculated binding energies and the rms radii.} 
\begin{tabular}{|c|c|c|c|c|c|}
${}^A$Z & $\hbar \omega$ [Mev] & - E$_{0}$ [MeV] 
& - E$_{exp}$ [MeV] & r${}_{rms}\, $[fm] & 1.2$A^{1/3}\, $[fm] \\
\hline
$^{6}$Li      & 19.47 & 30.1 & 30.5 & 2.41 & 2.18 \\  
$^{7}$Li      & 19.71 & 38.0 & 37.7 & 2.45 & 2.30 \\ 
$^{8}$Li      & 16.44 & 36.1 & 39.7 & 2.57 & 2.40 \\ 
$^{9}$Li      & 16.21 & 42.4 & 43.8 & 2.63 & 2.50 \\ 
$^{11}$Li     & 13.68 & 48.1 & 44.1 & 2.76 & 2.67 \\ 
$^{11}$C      & 16.01 & 72.3 & 70.4 & 2.61 & 2.67 
\end{tabular}
\end{table}

\bigskip 

\begin{table}
\caption{Parameters and rms radii for each shell, in fm. 
$r_o$ stands for the parameters of the outer/halo neutrons. 
$^{11}$Li$_{p}$ has four of them close to the p-shell; 
for $^{11}$Li$_{h}$ two neutrons are on the p-shell, and 
another two are loosely bound halos.}
\begin{tabular}{|c|c|c|c|c|c|c|c|}  
${}^A$Z       & $r_A$ & $r_p$ & $r_o$ & 
$\sqrt{<r^2>_s}$ & $\sqrt{<r^2>_p}$ & $\sqrt{<r^2>_o}$ 
& $\sqrt{<r^2>_{tot}}$ \\
\hline 
$^{6}$Li         & 1.46 & 1.52 &      & 1.84 & 2.50 & & 2.09 \\ 
$^{11}$Li$_{p}$  & 1.74 & 1.93 & 2.77 & 2.18 & 2.95 & 4.48 & 3.14 \\
$^{11}$Li$_{h}$  & 1.74 & 1.93 & 1.93 & 2.18 & 2.95 & 2.95 & 3.14 \\ 
 halo            &      &      & 3.28 &      &      & 5.23 &      \\
\hline 
$^{11}$Be        & 1.64 & 1.75 & 2.63 & 2.05 & 2.81 & 4.26 & 2.86 \\ 
\hline 
$^{11}$C         & 1.60 & 1.71 &      & 2.01 & 2.76 &      & 2.52 \\ 
$^{12}$C         & 1.61 & 1.77 &      & 2.02 & 2.86 &      & 2.61 \\
$^{12}$C {\small Eq. (\ref{hoho})} 
& 2.0 & 1.70 &      & 2.45 & 2.68 &      & 2.61 
\end{tabular}
\end{table}

\begin{table}
\caption{Experimental and calculated total reaction cross section 
$\sigma_R$ and inclusive cross section $\sigma_{^{11}C}$ for 
the reaction $^{12}$C + C, 1000 MeV/A. 
$\sigma_B$ and $\sigma_N$ are the widths of the broad 
and narrow momentum components. Cross sections are in mb.}
\begin{tabular}{|r|c|c|c|c|}
& $\sigma_B$  & $\sigma_N$  & $\sigma_R$ & $\sigma_{^{11}C}$ \\
\hline 
Experiment       & 105 &    & 930 & 47 \\
Ref. \cite{Har94}& 96  & 36 & 929 & 61 \\
$n(R_{0 K})$     & 87  & 53 & 943 & 65 \\
$n(\chi)$        & 96  & 40 & 937 & 63 \\
n(Eq.\ref{hoho}) & 113 & 45 & 929 & 48 
\end{tabular}
\end{table}

\begin{table}
\caption{Comparison of transverse momentum widths for $^{9}$Li fragments, 
experimental and calculated total reaction cross section 
and inclusive cross sections for the reaction $^{11}$Li + C, 790 MeV/A. 
$\sigma_B$ and $\sigma_N$ are the widths of the broad and 
narrow momentum components. $B/N$ is the ratio of their areas. 
Cross sections are in mb.}
\begin{tabular}{|r|c|c|c|c|c|c|c|}  
& $\sigma_B$  & $\sigma_N$  & $B/N$ & $\sigma_R$ & $\sigma_{^{9}Li}$
& $\sigma_{^{8}Li}$ & $\sigma_{^{8}He}$ \\
\hline 
Experiment            & 80 & 21 & 1.5  & 1042 & 213 & 62 & 26 \\
Ref. \cite{Har94}     & 80 & 21 & 1.22 & 1036 & 180 & 54 & 30 \\
n($^{11}$Li$_{h}$) & 80 & 28 & 1.65 & 994 & 147 & 48 & 30 \\
n($^{11}$Li$_{p}$) & 82 & 34 & 1.75 & 978 & 147 & 37 & 25 
\end{tabular}
\end{table}

\begin{table}
\caption{Comparison of $^{11}$Li, $^{11}$Be and $^{11}$C projectiles 
on a standard $^{12}$C target. All cross-sections are in mb.}
\begin{tabular}{|c|c|c|c|}
${}^{A}$Z & $\sigma_R$  & $\sigma_R$(exp) & Ref. \cite{mer} \\
\hline 
$^{11}$Li & 994 & 1042 & \\
$^{11}$Be & 946 & 942 & \\
$^{11}$C  & 884 &     & 800.6 
\end{tabular}
\end{table}

\end{document}